\documentclass[twocolumn,showpacs,preprintnumbers,
               amsmath,amssymb,epsfig]{revtex4}

\usepackage{epsfig}
\begin{document}

\def\omunit{{(km s$^{-1}$)/kpc}}
\def\gtrsim{ \lower .75ex \hbox{$\sim$} \llap{\raise .27ex \hbox{$>$}} }
\def\lesssim{ \lower .75ex \hbox{$\sim$} \llap{\raise .27ex \hbox{$<$}} }

\preprint{CITA-2001-76~~~PRL 89, 051102}

 \title{Cosmic Ray Diffusion from the Galactic Spiral Arms, \\ Iron
        Meteorites,  and a possible climatic connection?}
 \author{Nir J. Shaviv}
 \affiliation{
Canadian Institute for Theoretical Astrophysics, University of
Toronto\\ 60 St.~George Str., Toronto, ON M5S 3H8, Canada, and \\
Racah Institute of Physics, Hebrew University, Jerusalem 91904, Israel
}
\begin{abstract}

We construct a Galactic cosmic ray (CR) diffusion model while considering
that CR sources reside predominantly in the Galactic spiral arms. We find
that the CR flux (CRF) reaching the solar system should periodically
increase each crossing of a Galactic spiral arm. We search for this signal
in the CR exposure age record of Iron meteorites and confirm this
prediction.  We then check the hypothesis that climate, and in particular
the temperature, is affected by the CRF to the extent that glaciations can
be induced or completely hindered by possible climatic variations. We find
that although the geological evidence for the occurrence of IAEs in the
past Eon is not unequivocal, it appears to have a nontrivial correlation
with the spiral arm crossings---agreeing in period and phase. Thus, a
better timing study of glaciations could either confirm this result as an
explanation to the occurrence of IAEs or refute a CRF climatic connection. 

\end{abstract}


\maketitle



With the possible exception of extremely high energies, cosmic rays (CRs)
are believed to originate from supernova (SN) remnants
\cite{Longair,Berezinskii}. Moreover, most SNe in spiral galaxies like our
own are those which originate from massive stars, thus, they predominantly
reside in the spiral arms, where most massive stars are born and shortly
thereafter explode as SNe \cite{Drag1999}. Indeed, high contrasts in the
non-thermal radio emission are observed between the spiral arms and disks
of external galaxies. Assuming equipartition between the CR energy density
and the magnetic field, a CR density contrast can be inferred. It can have
a lower limit of 5 in some cases \cite{Duric1986}. 

Thus, while the Sun is crossing the Galactic spiral arms, the CRF is
expected to be higher.  To estimate the CRF variation, we construct a
simple diffusion model which considers that the CR sources reside in the
Galactic spiral arms. We expand the basic CR diffusion models (e.g.,
ref.~\cite{Berezinskii}) to include a source distribution located in the
Galactic spiral arms. Namely, we replace a homogeneous disk with an arm
geometry as given by Taylor \& Cordes \cite{Taylor1993}, and solve the
time dependent diffusion problem. To take into account the ``Orion spur''
\cite{Georgelin1976}, in which the Sun currently resides, we add an arm
``segment'' at our present location. Since the density of HII regions in
this spur is roughly half of the density in the real nearby arms
\cite{Georgelin1976}, we assume it to have half the typical CR sources as
the main arms. We integrate the CR sources assuming a diffusion
coefficient of $D= 10^{28}$cm$^{2}/$sec, which is a typical value obtained
in diffusion models for the CRs
\cite{Berezinskii,Webber1998,Lisenfeld1996}. We also assume a halo
half-width of 2kpc, which again is a typical value obtained in diffusion
models \cite{Berezinskii}, but more importantly, we reproduce with it the
$^{10}$Be survival fraction \cite{Lukasiak1994}. Thus, the only free
parameter in the model is the angular velocity $\Delta \Omega \equiv
\Omega_\odot-\Omega_p$ around the Galaxy of the solar system {\em
relative} to the Spiral arm pattern speed, which is later adopted using
observations. Results of the model are depicted in fig.~1. For the nominal
values chosen in our diffusion model and the particular pattern speed
which will soon be shown to fit various data, the expected CRF changes
from about 25\% of the current day CRF to about 135\%. Moreover, the
average CRF obtained in units of today's CRF is 76\%. This is consistent
with measurements showing that the average CRF over the period 150-700 Myr
before present (BP), was about 28\% lower than the current day CRF
\cite{Lavielle1999}. 

Interestingly, the temporal behavior is both skewed and lagging after the
spiral arm passages. The lag arises because the spiral arms are defined
through the free electron distribution. However the CRs are emitted from
SNe which on average occur roughly 15 Myr after the average ionizing
photons are emitted. The skewness arises because it takes time for the CRs
to diffuse after they are emitted. As a result, before the region of a
given star reaches an arm, the CR density is low since no CRs were
recently injected in that region and the sole flux is of CRs that succeed
to diffuse to the region from large distances.  After the region crosses
the spiral arm, the CR density is larger since locally there was a recent
injection of new CRs which only slowly disperse. This typically introduces
a 10 Myr lag in the flux, totaling about 25 Myr with the SN delay. This
lag is actually observed in the synchrotron emission from M51, which shows
a peaked emission trailing the spiral arms \cite{Longair}. 

The spiral pattern speed of the Milky Way has not yet been reasonably
determined through astronomical observations. Nevertheless, a survey of
the literature reveals that almost all observational determinations
cluster either around $\Delta \Omega \approx 9$ to 13 \omunit
\cite{list13} or around $\Delta \Omega \approx 2$ to 5 \omunit
\cite{list5}.  In fact, one analysis \cite{Palous1977} revealed that both
$\Delta \Omega$ = 5 or 11.5 \omunit ~fit the data. However, if the
spiral arms are a density wave \cite{Lin1964}, as is commonly believed
\cite{spiral}, then the observations of the 4-arm spiral structure in HI
outside the Galactic solar orbit \cite{NewMW} severely constrain the
pattern speed to $\Delta \Omega~\gtrsim~9.1 \pm 2.4$ \omunit,
since the four arm density wave spiral cannot extend beyond the outer 4 to
1 Lindblad resonance \cite{bigpaper}. We therefore expect the spiral
pattern speed obtained to coincide with one of the two aforementioned
ranges, with a strong theoretical argumentation favoring the first range. 

To validate the above prediction, that the CRF varied periodically, we
require a direct ``historic'' record from which the actual time dependence
of the CRF can be extracted.  To find this record, we take a compilation
of 74 Iron meteorites which were $^{41}$K/$^{40}$K exposure dated
\cite{Voshage}. CRF exposure dating (which measures the duration a given
meteorite was exposed to CRs) assumes that the CRF history was constant,
such that a linear change in the integrated flux corresponds to a linear
change in age. However, if the CRF is variable, the apparent exposure age
will be distorted. Long periods during which the CRF is low would
correspond to slow increases in the exposure age. Consequently, Fe
meteorites with real ages within this low CRF period would cluster
together since they will not have significantly different integrated
exposures. Periods with higher CRFs will have the opposite effect and
spread apart the exposure ages of meteorites. To avoid real clustering in
the data (due to one parent body generating many meteorites), we remove
all occurrences of Fe meteorites of the same classification that are
separated by less than 100 Myr and replace them by the average. This
leaves us with 42 meteorites. 

From inspection of fig.~1, it appears that the meteorites cluster with a
period of $143 \pm 10$~Myr, or equivalently, $\left|\Delta \Omega \right|
= 11.0 \pm 0.8$ \omunit, which falls within the preferred range for the
spiral arm pattern speed. If we fold the CR exposure ages over this
period, we obtain the histogram in fig.~2.  A K-S test yields a
probability of 1.2\% for generating this non-uniform signal from a uniform
distribution. Moreover, fig.~2 also describes the prediction from the CR
diffusion model. We see that the clustering is not in phase with the
spiral arm crossing, but is with the correct phase and shape predicted by
the CR model using the above pattern speed. A K-S test yields a 90\%
probability for generating it from the CR model distribution. Thus, we
safely conclude that spiral arm passages modulate the CRF with a $\sim
143$ Myr period. 

In 1959, Ney \cite{Ney1959} suggested that the Galactic CR flux (CRF)
reaching Earth could be affecting the climate since the CRF governs the
ionization of the lower atmosphere, to which the climate may in principle
be sensitive to. If this hypothesis is correct, we may be able to see a
correlation between the observed long term CRF variability and the climate
record on Earth. 

Interestingly, the CRF reaching Earth is also variable because of its
interaction with the variable solar wind. Thus, solar activity variations
will too have climatic effects {\em if the CRF affects the climate} (e.g.,
\cite{Soon2000}). Under the assumption that it does affect climate, we can
estimate how large an effect can a possible CRF-temperature relation be.
This can be derived from the fact that the best fit to the global warming
in the past 120 years is obtained if somewhat less then half is attributed
to anthropogenic greenhouse gases and somewhat more than half to the
increased activity of the sun \cite{Soon1996,Beer2000}. Thus, between
about 1940 and 1970, the global temperature, which decreased by
$0.15^\circ$K, is best explained as $-0.2^\circ$K attributed to the
reduced solar activity and $+0.05^\circ$K to greenhouse gases
\cite{Soon1996,Beer2000}. A global CRF climate effect is presumably more
likey to arise from CRs that can reach the troposphere and equatorial
latitudes. Thus, it is reasonable to assume that a possible effect would
arise from CRs that have high rigidities ($\gtrsim$10-15 GeV/nucleon). We
therefore normalize the low geomagnetic data from Haleakala, Hawaii and
Huancayo, Peru to the higher geomagnetic data of Cli max, Colorado
\cite{Bazil2000} that was measured over a longer period (e.g.,
\cite{Sven1998}). We find that the $-0.2^\circ$K cooling correlated with a
1.5\% increase in the high rigidity CRF. Thus, changing the CRF by $\pm
1\%$ would correspond to a global change of $\mp0.13^\circ$K, {\em on
condition that CRs are indeed the link relating solar activity to the
climate}. 

For the nominal values chosen in our diffusion model, the expected CRF
changes from about 25\% of the current day CRF to about 135\%.  This
corresponds to a temperature change of about $+10^{\circ}$K to
$-5^{\circ}$K, relative to today's temperature.  This range is sufficient
to markably help or hinder Earth from entering an IAE. 

\begin{figure}[tbh]
\vskip -0.1cm
\center{
\epsfig{file=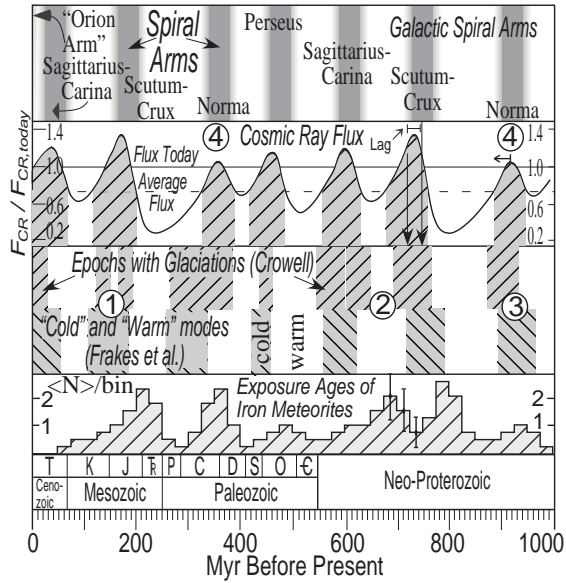,width=2.9in}
\vskip -0.25cm
 }
\caption{ The past Eon. Panel A describes past Galactic spiral arms
crossings assuming $\Delta \Omega=10.9$ \omunit.  Panel B
describes the CRF reaching the solar system using the CR diffusion model. 
Note that the CRF lags behind spiral arm crossings. This is portrayed by
the hatched regions, which qualitatively show the predicted occurrences of
IAEs if the CRF required to trigger them is the average flux.  Arrows mark
the middle of the spiral crossing and the expected mid-glaciation point. 
Panel C qualitatively describes the geologically recorded IAEs---its top
half, as summarized by Crowell \cite{Crowell1999}, who concluded that the
current evidence is insufficient to support a claim of periodicity, while
the bottom half, as summarized by Frakes et al. \cite{Frakes} who claim a
periodicity exists. By fine tuning the observed pattern speed of the arms
to best fit the IAEs, an intriguing correlation appears between the IAEs
and their prediction.  Note that the correlation need not be absolute
since additional factors may affect the climate.  Other factors that
should be considered and noted in the graph are: (1) The mid-Mesozoic
glaciations are significantly less extensive as others. (2) It is unclear
to what extent was the period around 700 Myr BP warmer than the IAEs
before or after. (3) The first IAE of the Neo-Proterozoic (if indeed
distinct) is very uncertain. (4) Since Norma's crossing is an
extrapolation from smaller galactic radii, its location is uncertain. If
the arm's structure at smaller radii is indeed different \cite{bigpaper},
its preferred location will lag by about 20 Myr. Panel D is a 1-2-1
averaged histogram of the $^{41}$K/$^{40}$K exposure ages of Fe
Meteorites, which are predicted to cluster around the CRF minima. The
cluster-IAE correlation further suggests an extra-terrestrial trigger for
the glaciations. } 
\end{figure} 

 \begin{figure}[tbh]
\center{\epsfig{file=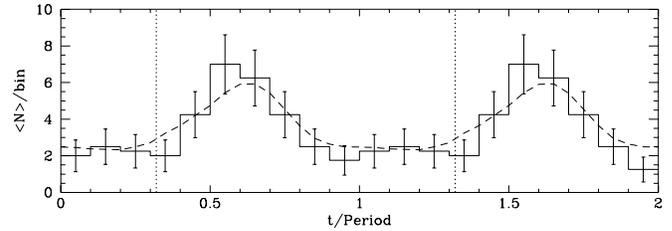,width=1.2in,angle=-90}}
\caption{A folded 1-2-1 averaged histogram (over the IAE periodicity) of
the Fe Meteorites' exposure ages. A statistically significant signal is
obtained for a period of $143\pm 10$ Myr. The dashed line is the CR
diffusion model prediction. We find that the predicted phase of the
cluster peaks, agrees with the actual exposure age clustering. The
clustering peaks 100 Myr after the spiral crossing (dotted lines),
implying that it is unlikely to be a {\em real} age clustering peak, which
should be related directly to the spiral crossing (e.g., by Oort cloud
comets being perturbed during arm crossing, that subsequently break
asteroids more frequently, and that somehow affect the terrestrial
climate).}
\end{figure}

Extensive summaries of IAEs on Earth can be found in Crowell
\cite{Crowell1999} and Frakes et al. \cite{Frakes}.  Those of the past Eon
are summarized in fig.~1. The nature of some of the IAEs is well
understood while others are sketchy in detail.  The main uncertainties are
noted in fig.~1. For example, it is unclear to what extent can the milder
mid-Mesozoic glaciations be placed on the same footing as other IAEs, nor
is it clear to what extent can the period around 700 Myr BP be called a
warm period since glaciations were present, though probably not to the
same extent as the periods before or after. Thus, Crowell
\cite{Crowell1999} concludes that the evidence is insufficient to claim a
periodicity. On the other hand, Williams \cite{Williams1975} claimed that
a periodicity may be present. This was significantly elaborated upon by
Frakes et al. \cite{Frakes}. 

Comparison between the CRF and the glaciations in the past 1 Gyr shows a
compelling correlation (fig.~1).  To quantify this correlation, we perform
a $\chi^2$ analysis. {\em To be conservative}, we do so with the Crowell
data which is less regular. Also, we do not consider the possible IAE
around 900 Myr, though it does correlate with a spiral arm crossing. For a
given pattern speed, we predict the location of the spiral arms using the
model. We find that a minimum is obtained for $\Delta \Omega =
10.9\pm 0.25$ \omunit, with $\chi^2_{min}=1.1$ per degree of freedom (of
which there are 5=6-1).  We also repeat the analysis when we neglect the
lag and again when we assume that the spiral arms are separated by
$90^\circ$ (as opposed to the somewhat asymmetric location obtained by
Taylor and Cordes \cite{Taylor1993}). Both assumptions degrade the fit
($\chi^2_{min}=2.9$ with no lag, and $\chi^2_{min}=2.1$ with a symmetric
arm location). Thus, the latter analysis assures that IAEs are more likely
to be related to the spiral arms and not a more periodic phenomena, while
the former helps assure that the CRs are more likely to be the cause,
since they are predicted (and observed) to be lagged. 

The previous analysis shows that to within the limitation of the
uncertainties in the IAEs, the predictions of the CR diffusion model and
the actual occurrences of IAE are consistent.  To understand the
significance of the result, we should also ask the question what is the
probability that a random distribution of IAEs could generate a $\chi^2$
result which is as small as previously obtained. To do so, glaciation
epochs where randomly chosen. To mimic the effect that nearby glaciations
might appear as one epoch, we bunch together glaciations that are
separated by less than 60 Myrs (which is roughly the smallest separation
between observed glaciations epochs). The fraction of random
configurations that surpass the $\chi^2$ obtained for the best fit found
before is of order 0.1\% for {\em any} pattern speed. (If glaciations are
not bunched, the fraction is about 100 times smaller, while it is about 5
times larger if the criterion for bunching is a separation of 100 Myrs or
less). The fraction becomes roughly $6\times 10^{-5}$ (or a 4-$\sigma$
fluctuation), to coincidentally fit the actual period seen in the Iron
meteorites. 

Last, before 1 Gyr BP, there are no indications for any IAEs, except for
periods around 2 - 2.5 Gyr BP (Huronian) and 3 Gyr BP (late
Archean)\cite{Crowell1999}.  This too has a good explanation within the
picture presented. Different estimates to the Star formation rate (SFR) in
the Milky Way (and therefore also to the CR production) point to a peak
around 300 Myr BP, a significant dip between 1 and 2 Gyr BP (about a third
of today's SFR) and a most significant peak at 2-3 Gyr BP (about twice as
today's SFR)\cite{Scalo1987,RochaPinto2000}. This would imply that at 300
Myr BP, a more prominent IAE should have occurred---explaining the large
extent of the Carboniferous-Permian IAE. Between 1 and 2 Gyr BP, there
should have been no glaciations and indeed none were seen. Last, IAEs
should have also occurred 2 to 3 Gyr BP, which explains the Huronian and
late-Archean IAEs. 

To conclude, by considering that most CR sources reside in the Galactic
spiral arms, we predict a variable CRF. A record of this signal was indeed
found in Iron meteorites, and it nicely agrees with the observations of
the Galactic spiral arm pattern speed.  Next, if the apparent solar
activity climate correlation is real and arises from modulation of the
galactic CRF reaching Earth, then typical variations of up to ${\cal
O}(10^\circ {\rm K})$ could be expected from the variable CRF. Each spiral
arm crossing, the average global temperature should reduce enough to
trigger an IAE.  The record of IAEs on Earth is fully consistent with the
predicted and observed CRF variation---both in period and in phase.
Moreover, the fit improves when the predicted lag in the IAEs after each
crossing is included and when the actual asymmetric location of the arms
is considered. Moreover, a random mechanism to generate the IAEs is
excluded. Nevertheless, one should bear in mind that the weakest link
still remains the glaciological record with its uncertainties. That is,
more research on the timing and extent of glaciations is required. 

The last agreement is between the Eon time scale star formation activity
of the Milky Way and presence or complete absence of IAEs. Here, a more
detailed research on the SFR activity would be useful to strengthen (or
perhaps refute) the long term correlation.

If the apparent correlation between observed CRF variations and climate on
Earth is not simply a remakable coincidence, an unavoidable question is
what is the physical mechanism behind the CRF/temperature relation?
Currently, there is no single {\em undisputed} mechanism through which
cosmic rays can affect the climate. There are however several
observational indications that such a relation could exist. For example,
Forbush events during which the CRF suddenly drops on a time scale of days
were found to correlate with the amount of ``storminess" as encapsulated
by the vorticity area index \cite{Tinsley1991}, or a concurrent drop in
the cloud cover \cite{Pudovkin}. There were also claims that the galactic
CRF, which his modulated by the solar cycle and slightly lags behind it,
correlates with the low altitude cloud cover variations
\cite{Sven1998,Marsh2000}. Clearly, an in depth study on the possible
climatic effects of cosmic-rays is imperative. 

The author is particularly grateful to Peter Ulmschneider for the
stimulating discussions which led to the development of this
idea. The author also wishes to thank Norm Murray, Chris Thompson,
and Joe Weingartner for their very helpful comments and suggestions.

\def\listlow{The first range of results for
$\Delta \Omega$ includes $\sim 11.5$ \omunit, C.~C. {Lin},
C.~{Yuan}, F.~H. {Shu}, {\em \apj} {\bf 155}, 721 (1969); $\sim 11.5$,
C.~Yuan, {\em \apj}{\bf 158}, 871, (1969); $\sim 11.5$, C.~Yuan, {\em
\apj}{\bf 158}, 889, (1969); $13.5\pm1.5$, M.~A.~{Gordon}, {\em \apj}{\bf
222}, 100 (1978); $\sim 11.5$ \cite{Palous1977}; $9-13$, E.~M.~{Grivnev},
{\em Sov. Astron. Lett.} {\bf 9}, 287 (1983); and $7.5-11.5$,
G.~R.~{Ivanov}, {\em Pis ma Astron.  Zhurnal} {\bf 9}, 200 (1983)}

\def\listhigh{The second range of results for
$\Delta \Omega$ includes $2.5 \pm 1.5$ \omunit, M.~{Creze}, M.~O.
{Mennessier}, {\em \aap}{\bf 27}, 281, (1973); $\sim 5$,
\cite{Palous1977}; $\sim 5$, A.~H.~Nelson, T.~Matsuda, {\em \mnras}{\bf
179}, 663 (1977); $1.4\pm3.6$, I.~N.~{Mishurov}, E.~D.~{Pavlovskaia},
A.~A.~{Suchkov}, {\em Astron. Zhurnal} {\bf 56}, 268 (1979); $2-4$,
E.~M.~{Grivnev}, {\em Sov. Astron. Lett.} {\bf 7}, 303 (1981); and $2.3
\pm 1$, L.~H.~{Amaral}, J.~R.~D.~{Lepine}, {\em \mnras}{\bf 286}, 885
(1997)}

\def\HEbook{M. S. Longair, {\em High Energy Astrophysics}, 2$^{\rm nd}$
ed., vol. 2 (Cambridge Univ. Press, Cambridge, 1994) }

\def\CRbook{V.~S. Berezinski\u{\i}, S. V. Bulanov, V. A. Dogiel, V. L.
Ginzburg, V. S. Ptuskin, {\em Astrophysics of Cosmic Rays},
(North-Holland, Amsterdam, 1990) }

\def\GalDynBook{J. Binney, S. Tremaine, {\em Galactic Dynamics},
(Princeton Univ. Press, Princeton, 1988)  }

\newcommand{\mnras}{{Mon. Not. Roy. Astr. Soc.~}}
\newcommand{\aap}{{Astron.~Astrophy.~}}
 


\end{document}